\documentclass[preprint]{vgtc}               

\graphicspath{{figures/}} 

\usepackage{times}                     

\usepackage{mathptmx}                  

\usepackage{booktabs}
\usepackage{tabularx}
\usepackage{colortbl}
\usepackage{multibib}

\newcites{app}{Appendix References}

\onlineid{0}

\vgtccategory{Research}

\vgtcinsertpkg

\title{Exploring Text Classification Models with Sparse Autoencoders}

\author{Daniel Kerrigan\thanks{e-mail: daniel.kerrigan2@capitalone.com}\\ %
        \scriptsize Capital One %
\and Brian Barr\thanks{e-mail: brian.barr@capitalone.com}\\ %
     \scriptsize Capital One %
\and Enrico Bertini\thanks{e-mail: e.bertini@northeastern.edu}\\ %
     \scriptsize Northeastern University}

\teaser{
  \centering
  \includegraphics[width=\linewidth]{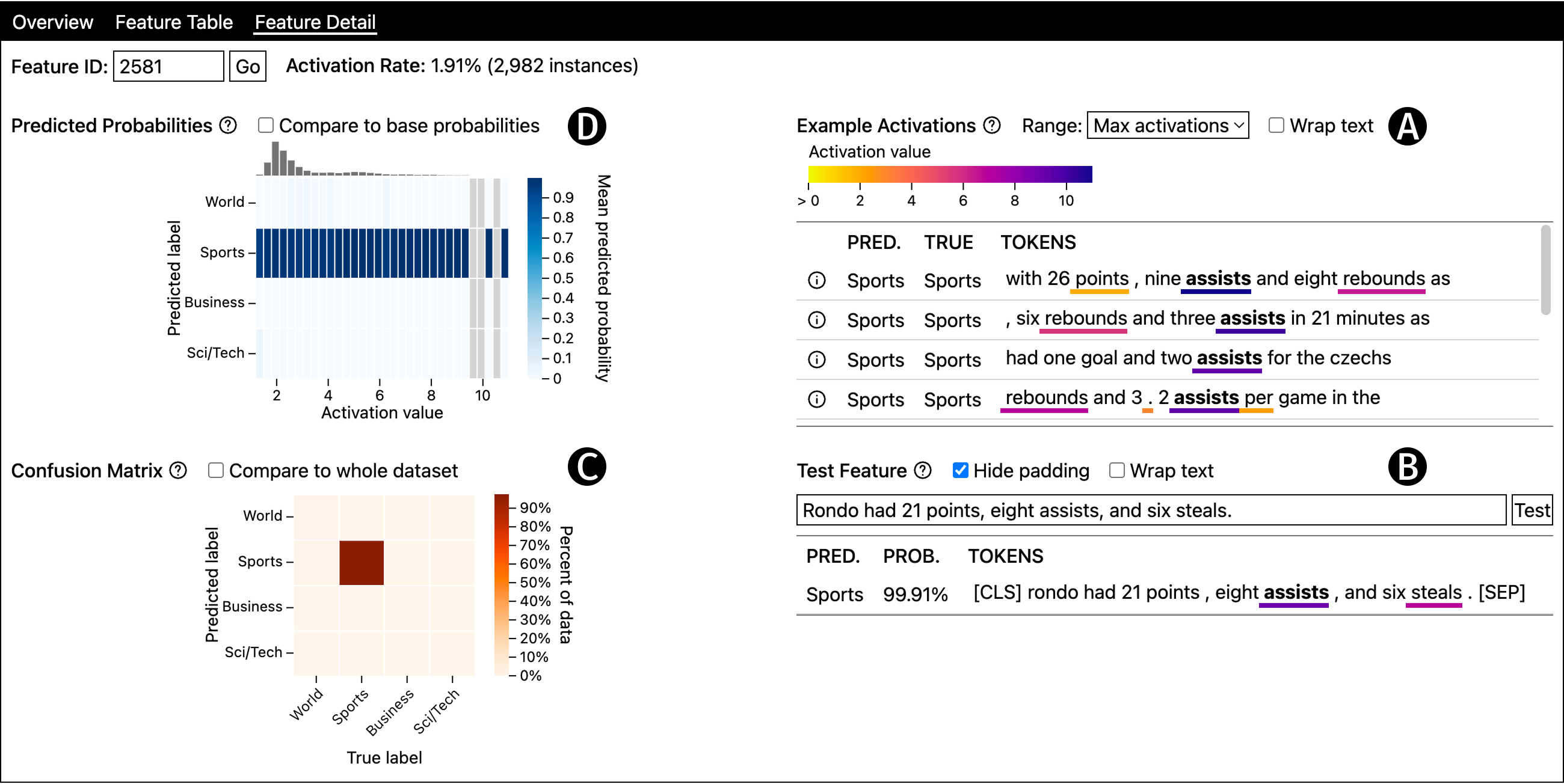}
  \caption{%
        The Feature Detail tab in SAEfarer for a model that classifies the topic of news articles.
        This feature activates on tokens relating to sports statistics.
        A) The example activations shows snippets from instances that activate the feature.
        B) The user can enter their own piece of text to see if it activates the feature.
        C) A confusion matrix calculated from the instances that activate the feature.
        D) The relationship between the feature's activation value and the model's predicted probabilities for each class.
    }
    \label{fig:saefarer_detail_sports}
}

\abstract{
As language models (LMs) rise in prominence, there is interest in making them more transparent in order to better understand their internal behavior. Recent interpretability work has focused on using sparse autoencoders (SAEs) to break down neuron activations at a given layer in the LM into human-understandable features, where each feature represents a concept that the model has learned. In this paper, we share work on using SAEs to analyze the behavior of text classification LMs. We present techniques for exploring the relationships between the SAE's features and the model's predictions and errors. We integrate these techniques into SAEfarer, a tool for analyzing concepts learned by text classification LMs. We assess SAEfarer in an expert pilot evaluation with five Ph.D. students.
} 


\begin{document}

\firstsection{Introduction}

\maketitle

When interpreting and explaining the behavior of machine learning (ML) models trained on tabular data, much of the focus is on analyzing the relationships between the models' features and predictions (e.g.,~\cite{friedman2001GreedyFunctionApproximation, lundberg2017UnifiedApproachInterpreting}).
One difficulty in analyzing the behavior of models trained on unstructured data like images or text is that they lack interpretable features in the traditional sense.
Individual pixels or tokens that form the inputs to the model may be too granular to promote a higher-level conceptual understanding of the model from analyzing them~\cite{kim2018InterpretabilityFeatureAttribution}.
To address this, researchers have developed techniques to extract concepts learned by image and language models (LMs) (e.g., \cite{ghorbani2019AutomaticConceptbasedExplanations,bricken2023MonosemanticityDecomposingLanguage}).
These approaches can be viewed as ways to identify higher-level ``features'' of the models.

Visual analytics research on concept-level analysis of neural networks has focused primarily on image models~\cite{park2022NeuroCartographyScalableAutomatic, huang2023ConceptExplainerInteractiveExplanation, zhang2024SlicingChattingRefining, zhao2022HumanintheloopExtractionInterpretable} and generative LMs~\cite{bricken2023MonosemanticityDecomposingLanguage, lin2023NeuronpediaInteractiveReference}.
In this work, we present a concept-based visual analytics approach for exploring the behavior of text classification LMs. 
Our approach relies on sparse autoencoders (SAEs) to identify features learned by the model~\cite{bricken2023MonosemanticityDecomposingLanguage, cunningham2023SparseAutoencodersFind}.

One challenge in working with SAEs is the large number of features that they extract.
When an ML practitioner is exploring the behavior of the model, it is impractical for them to analyze all of the identified features.
It is necessary to have ways to help guide the practitioner's exploration to surface features of particular interest.
When analyzing text classification LMs, of key interest are features that correlate with the model's predictions and errors.
Analyzing the features that correlate with the model's predictions can help the practitioner check if the model has learned intuitive concepts that are useful for classification.
Analyzing the features that correlate with the model's errors can help the practitioner identify patterns in the instances that the model makes mistakes on.

In this work, we first propose ways of ranking SAE features based on their relationships with the text classification model's predictions and errors.
Second, we present SAEfarer, an interactive visualization tool for exploring SAE features for text classification LMs.
Finally, we discuss an expert pilot evaluation with five Ph.D. students whose research relates to machine learning interpretability.

\section{Related Work}

\subsection{Identifying Concepts in Models}

TCAV by Kim et al.~\cite{kim2018InterpretabilityFeatureAttribution} identified the direction of predefined concepts in the activation space of a neural network and measured the relationship between those concepts and the model's predictions.
They demonstrated their approach on image models.
Ghorbani et al. extended this work with ACE~\cite{ghorbani2019AutomaticConceptbasedExplanations} in order to overcome the limitation of requiring predefined concepts.
Later work by Yeh et al.~\cite{yeh2020CompletenessawareConceptBasedExplanations} proposed an approach to automatically identify a set of concepts that sufficiently explain the model's behavior, which they demonstrated for both image and text classification models.

Bricken et al.~\cite{bricken2023MonosemanticityDecomposingLanguage} and Cunningham et al.~\cite{cunningham2023SparseAutoencodersFind} presented notable initial work on using SAEs to identify interpretable concepts learned by transformer-based LMs.
These initial works extracted concepts from small or toy models, but later work applied SAEs to larger LMs such as GPT-4~\cite{gao2024ScalingEvaluatingSparse}, Claude 3 Sonnet~\cite{templeton2024ScalingMonosemanticityExtracting}, and Gemma 2 and 3 ~\cite{lieberum2024GemmaScopeOpen, mcdougall2025GemmaScope2}.
Extracting concepts with SAEs is unsupervised and does not require labeled data or a human to predefine concepts.
Despite not using labels to identify concepts, SAEs are still a useful technique to apply with labeled data.

Movva et al.~\cite{movva2025SparseAutoencodersHypothesis} proposed a method called HypotheSAEs that uses SAEs to generate hypotheses about the relationship between a text dataset and a target variable.
Although both SAEfarer and HypotheSAEs use SAEs to analyze labeled text data, the works have different focuses.
SAEfarer is a visualization system for exploring the relationships between SAE features and a model's predictions and errors.
In contrast, HypotheSAEs is a method for generating natural language hypotheses about a labeled text dataset and does not focus on analyzing the behavior of a classification model.

\subsection{Concept Visualization Systems}

Visual analytics research in concept-level analysis of neural networks initially focused on image models.
Zhao et al.~\cite{zhao2022HumanintheloopExtractionInterpretable} proposed an active learning system called ConceptExtract for interactively identifying concepts.
With NeuroCartography, Park et al.~\cite{park2022NeuroCartographyScalableAutomatic} created a tool to identify clusters of neurons that activate for the same concept and to analyze the relationship between concepts.
Huang et al.~\cite{huang2023ConceptExplainerInteractiveExplanation} presented ConceptExplainer, a system for exploring concepts based on TCAV and ACE.
Zhang et al.~\cite{zhang2024SlicingChattingRefining} proposed ConceptSlicer, which uses concepts to identify subsets of the data where the model performs poorly.

For LMs, researchers have developed tools for analyzing features identified by SAEs.
Bricken et al.~\cite{bricken2023MonosemanticityDecomposingLanguage} is particularly influential to our work, as they developed, to our knowledge, the first visualization tool for ranking and exploring SAE features.
McDougall's SAE Visualizer~\cite{mcdougall2024SAEVisualizer} and the SAEDashboard by Decode Research~\cite{decoderesearch2024SAEDashboard} are open-source tools based on the work by Bricken et al.~\cite{bricken2023MonosemanticityDecomposingLanguage}.
Neuronpedia~\cite{lin2023NeuronpediaInteractiveReference} is a notable SAE exploration and visualization tool that provides additional functionality such as testing feature activations on custom text and using SAE features to steer the model's output.
We have taken inspiration from these works and have adopted functionality from them in SAEfarer.
However, with SAEfarer, our focus is on using SAEs to analyze text classification LMs, rather than generative LMs.
For classification LMs, these existing tools would not directly guide the user to features that are relevant to the classification task and would not show the features' relationships to the model's predictions.
SAE Semantic Explorer by Yan et al.~\cite{yan2025visualexplorationfeaturerelationships} is another SAE tool for language, but its focus is on analyzing the relationships between features for specific concepts that the user is interested in.

\section{Training and Analyzing SAEs}

\begin{figure}
    \centering
    \includegraphics[width=\linewidth]{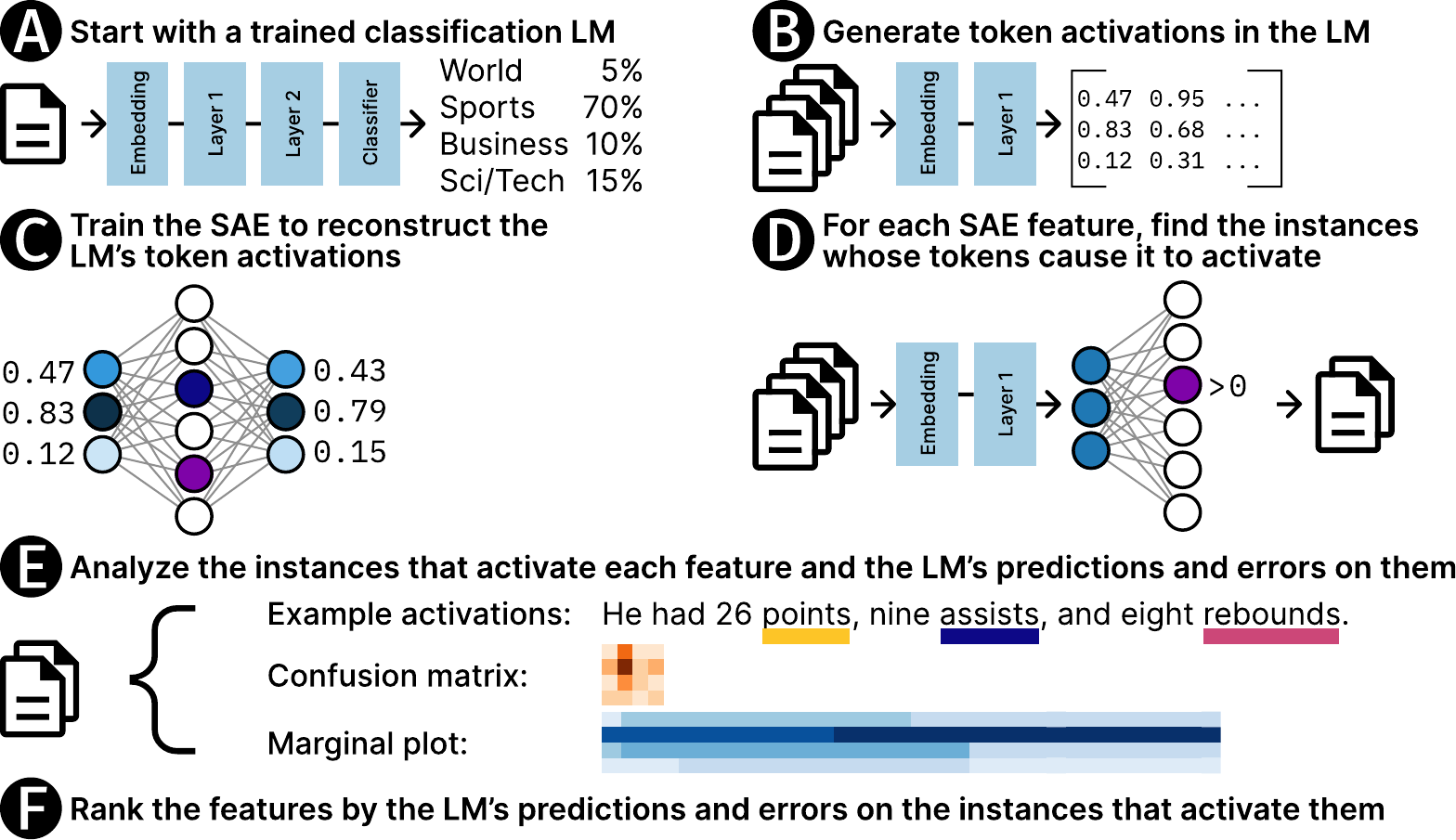}
    \caption[A summary of training and analyzing sparse autoencoders (SAEs) in SAEfarer.]{%
        Our approach to training and analyzing SAEs.
    }
  \label{fig:saefarer_approach}
\end{figure}

\Cref{fig:saefarer_approach} provides an overview of how we use SAEs in SAEfarer.
We begin with a trained classification LM~(\cref{fig:saefarer_approach}A).
As an example, we will use a model\footnote{\href{https://huggingface.co/Kyle1668/ag-news-19200-bert-base-uncased}{https://huggingface.co/Kyle1668/ag-news-19200-bert-base-uncased}} that was trained to classify the topic of articles from the AG News dataset~\cite{gulliAGsCorpusNews} as \textit{World}, \textit{Sports}, \textit{Business}, or \textit{Sci/Tech}.
To make a classification, an instance is split into a series of tokens before being fed into the model.
The model then outputs a score for each class and the class with the highest score is chosen as the predicted label for the instance.

The first step to training an SAE is to feed instances into the LM and generate activations for each token at a chosen layer in the LM (\cref{fig:saefarer_approach}B).
The activations for each token are represented as a high-dimensional vector.
For the model in our example, each activation vector has 768 dimensions.
An SAE is a neural network with one hidden layer that is trained to reconstruct these activations~(\cref{fig:saefarer_approach}C).
The hidden layer of the SAE is larger than the activation vectors.
In our example, we trained an SAE with 6144 hidden neurons.
Sparsity is enforced on the hidden layer of the SAE so that only a small number of neurons in the hidden layer can fire at once to reconstruct the activations for a given token.
In our case, we trained a k-sparse autoencoder~\cite{gao2024ScalingEvaluatingSparse} with k=8, meaning that only 8 of the 6144 hidden neurons can fire at the same time.
The SAE is trained to minimize the error between the input activations and the output reconstructed activations.

To analyze the trained SAE, we again feed instances through the LM to generate token activations~(\cref{fig:saefarer_approach}D).
We then feed those token activations into the SAE and observe the activations of the SAE's hidden-layer neurons.
In our example, for each token that is input to the LM, 8 of the SAE's hidden-layer neurons will have positive activation values.
We refer to the SAE's hidden-layer neurons as features.
For each feature in the SAE, we track which instances contain tokens that cause the feature to activate.

We then generate visualizations for each feature to help us understand when it activates and how it relates to the model's predictions and errors~(\cref{fig:saefarer_approach}E).
Ideally, each SAE feature represents a single, human-understandable concept.
The concept that a feature represents can be determined by analyzing example tokens that activate the feature and their surrounding contexts.
A good feature will primarily activate in contexts that are semantically similar.
A poor feature may activate on tokens relating to multiple distinct concepts or not have a clear pattern in its activations.
The goal of SAEfarer is to guide the user’s exploration of the SAE features and how they relate to the classification model’s behavior.
To help identify interesting features to explore, the last step in our approach is to rank the features based on a variety of metrics calculated from the model's predictions and errors on the instances that activate them~(\cref{fig:saefarer_approach}F).

\section{Feature Ranking}

One challenge with using SAEs to analyze the behavior of models is that they contain too many features to be able to exhaustively explore all of them.
It is necessary to identify and prioritize the most interesting features.
For text classification, we believe that it is particularly relevant to identify features that have strong relationships with the model's predictions and errors.
We propose an approach for ranking the features based on confusion matrices.

\newcommand{\graymidrule}{\arrayrulecolor[HTML]{D3D3D3}\specialrule{0.04pt}{0.32em}{0.32em}\arrayrulecolor{black}}
\begin{table}
    \centering
    \caption{%
    Example SAE feature ranking metrics.
    }
    \label{tab:saefarer_rankings}
    \small
    \begin{tabularx}{\columnwidth}{llX}
        \toprule

        Pred. label & True label & Ranking Metric\\

        \midrule

        \textit{World} & \textit{World} & \% of instances that activate the feature that are correctly classified as \textit{World}.\\

        \graymidrule

        \textit{Business} & \textit{Sci/Tech} & \% \ldots\ that are predicted as \textit{Business}, but are \textit{Sci/Tech}.\\

        \graymidrule

        \textit{Sports} & \texttt{Any}     & \% \ldots\ that are predicted to be \textit{Sports}.\\

        \graymidrule

        \texttt{Any}     & \textit{World} & \% \ldots\ that have a ground-truth label of \textit{World}.\\

        \graymidrule

        \textit{Business} & \texttt{Different}   & \% \ldots\ that are incorrectly predicted to be \textit{Business}.\\

        \graymidrule

        \texttt{Different}   & \textit{Sci/Tech} & \% \ldots\ that have a ground-truth label of \textit{Sci/Tech} and are incorrectly classified.\\

        \graymidrule

        \texttt{Any}     & \texttt{Different}   & \% \ldots\ that are incorrectly classified.\\

        \bottomrule
    \end{tabularx}
\end{table}

Each feature in the SAE is activated by a subset of instances in the dataset.
We consider an instance to activate a feature if at least one token in the instance activates the feature.
For each feature, we compute a confusion matrix for the instances that activate it.
We then rank the features based on metrics that we calculate from values in the confusion matrices, which we summarize in \cref{tab:saefarer_rankings}.

For ranking the features, the user selects a value for the predicted class label and the true class label.
For example, if the user selects the predicted label to be \textit{Business} and the true label to be \textit{Sci/Tech}, then this ranks the features based on the percentage of instances that activate the feature where the model predicts \textit{Business}, but the true label is \textit{Sci/Tech}.
This would identify features that have the strongest relationship with the model making that specific error.

In addition to selecting a specific class for the predicted or true label, the user can also select from two wildcard values.
The first wildcard is \texttt{Any}, which represents any class.
For example, if the user selects \textit{Sports} for the predicted label and \texttt{Any} for the true label, then this would rank the features based on the percentage of instances that activate the feature where the model predicts \textit{Sports}.
The second wildcard is \texttt{Different}, which represents any class except the other class that is selected.
For example, if the user selects \textit{Business} as the predicted label and \texttt{Different} as the true label, then \texttt{Different} represents \textit{World}, \textit{Sports}, and \textit{Sci/Tech}.
This configuration would rank the features based on the percentage of instances that activate the feature where the model incorrectly predicts the label to be \textit{Business}.
Finally, selecting \texttt{Any} as the predicted label and \texttt{Different} as the true label (or vice versa) would rank the features based on the percentage of instances that activate the feature where the model is incorrect.

\section{User Interface}
\label{sec:saefarer_ui}

SAEfarer is a tool for exploring the concepts learned by text classification LMs.
The primary goal is to enable the user to analyze the relationships between the features and the model's predictions and errors.
Our aim is to help users identify features of interest, understand the concepts that those features represent, and analyze the relationship that those features have with the model's behavior.
To do this, we integrated the feature rankings into an interactive widget for Jupyter notebooks~\cite{manz2024AnywidgetReusableWidgets, manz2024AnyNotebookServed}.
We organized the user interface into three tabs: Overview, Feature Detail, and Feature Table.

\subsection{Overview}

The Overview tab provides a high-level summary of the SAE and the model.
A series of tables provides information about the dataset, model, and SAE, such as the number of tokens in the dataset, the model's error rate, and the size of the SAE.
These tables are accompanied by two visualizations.
First, a confusion matrix shows the performance of the classification model.
Second, a histogram shows how often the features in the SAE activate.
We define the activation rate of a feature as the percentage of instances in the dataset where the feature activates on at least one token.

\subsection{Feature Detail}

The Feature Detail tab shows information about a single feature.
\Cref{fig:saefarer_detail_sports} provides an example of a feature that activates on descriptions of statistics from a sporting event.
At the top right (\cref{fig:saefarer_detail_sports}A), we display portions of instances that activated the feature.
By default, we show instances that caused the feature to have the highest activation values.
To visualize the snippets, we encode the activation value of the feature with a colored line beneath the token.
Under the example activations, the user can enter their own piece of text and check how it causes the feature to activate (\cref{fig:saefarer_detail_sports}B).

In the bottom left (\cref{fig:saefarer_detail_sports}C), we show a confusion matrix of the instances that activated the feature.
In this example, we can see that the feature has a strong relationship with the model accurately classifying \textit{Sports} articles.
In the top left (\cref{fig:saefarer_detail_sports}D), a marginal plot~\cite{apley2020VisualizingEffectsPredictor} shows the relationship between the feature's activation values and the model's predicted probabilities for each class.
For this plot, we bin the instances that activate this feature by their activation value.
If multiple tokens in the instance activate the feature, then we take the maximum value.
For each bin of instances, we compute the model's mean predicted probability for each class.
We then plot this as a heatmap.
For the feature in \cref{fig:saefarer_detail_sports}, we see that the model assigns a high probability to the \textit{Sports} class at all activation values.
Above this, we provide a histogram that shows the number of instances in each activation value bin.
The appendix contains additional example features.

\subsection{Feature Table}

\begin{figure*}
    \centering
    \includegraphics[width=\linewidth]{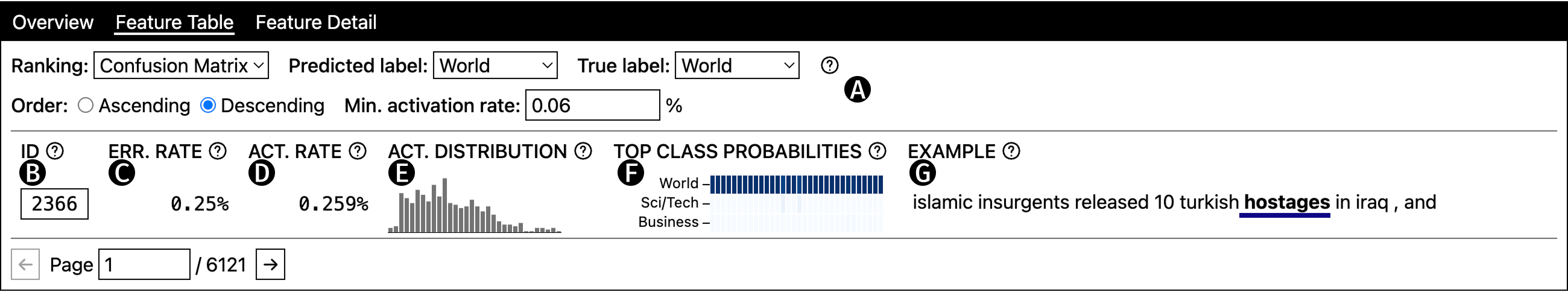}
    \caption{%
    The Feature Table tab.
    A) Controls for ranking and filtering the features.
    B-D) The feature's index, error rate, and activation rate.
    E) A histogram of the feature's activation values.
    F) The model's probabilities for the top classes.
    G) The top example that activates the feature.
    }
    \label{fig:saefarer_table}
\end{figure*}

The Feature Table tab enables the user to rank and filter the SAE features.
Its purpose is to help the user identify features that are worth analyzing in the Feature Detail tab.
\Cref{fig:saefarer_table} shows a condensed screenshot of the Feature Table tab.
At the top of the tab (\cref{fig:saefarer_table}A), there are controls that allow the user to adjust how they rank the features.
The main area of the tab contains a table in which each row represents one feature.
At the beginning of each row is a button that brings the user to the Feature Detail tab for the feature (\cref{fig:saefarer_table}B).
Next, we show the model's error rate on instances that activate the feature (\cref{fig:saefarer_table}C) and the feature's activation rate (\cref{fig:saefarer_table}D).
The remaining columns contain condensed versions of visualizations shown in the Feature Detail tab (\cref{fig:saefarer_table}E-G).

\section{Expert Pilot Evaluation}

We conducted an expert pilot evaluation to analyze how ML researchers use SAEfarer to explore a model.
We were interested in assessing the usability of the tool and how it could be improved.
Our evaluation was approved by the Northeastern University IRB (\#23-04-35) and all participants reviewed a participant information sheet.
Our participants were five Ph.D. students whose research relates to ML interpretability or explainability.
They all indicated that they were familiar with the use of SAEs for interpreting LMs.
We compensated each with a \$50 Amazon gift card.

\subsection{Protocol}

We met with each participant through a video call for two hours.
Each meeting began with a presentation where we gave an overview of the study, reviewed SAEs, and provided a tutorial on how to use SAEfarer.
We then gave the participant practice using SAEfarer.
We provided them with a Jupyter notebook to analyze an SAE that we trained on an example model and dataset.
We then asked them to answer a set of questions using the tool.
These questions gave them practice with SAEfarer and allowed us to check that they could correctly use it and interpret the visualizations.
We then gave the participant a few minutes to explore the model.

Next, we gave the participant a notebook that ran SAEfarer with an SAE trained on a model that classifies articles from the AG News dataset.
We gave the participant 30 minutes to explore the behavior of this model.
We instructed them to focus on analyzing the relationship between the SAE features and the model's predictions and errors.
We also provided them with the following example questions.
What features correlate with the model predicting the \textit{Sports} label?
What features correlate with the model making mistakes?
Are there any surprising or unintuitive features?
We chose an open-ended evaluation to allow participants to have different exploration strategies and to use the tool as they saw fit.
After the 30 minutes, we reviewed aspects of their exploration with them before moving on to an interview where we elicited feedback.

\subsection{Results}

The meetings with the participants were recorded and transcribed by Zoom.
We conducted thematic analysis~\cite{braun2006UsingThematicAnalysis} on the interview transcripts with the goal of understanding the strengths and weaknesses of SAEfarer and how it might be improved.
Overall, participants liked SAEfarer, but pointed out several ways to improve it.
Below we describe their feedback, organized by topic.

\textbf{Confusion matrix rankings.}
Participants overall found the confusion matrix rankings to be helpful and clear to use.
However, there is room for improvement to make it more obvious how to rank the features by error rate.
Likewise, it was not entirely clear that ranking the features in ascending order by this metric would identify the features on whose activating instances the model had the highest accuracy.
Allowing the user to adjust the rankings through the column headers could improve this.

\textbf{Providing more context.}
In the example activations for a feature~(\cref{fig:saefarer_detail_sports}A), we show the token that has the highest activation in the instance and a few tokens surrounding it.
One common request that we received was to be able to look at the entire instance and not just the provided snippet.
This would be useful in a few situations.
For example, some participants came across features with high error rates, where, based on the example activations,  they noted that it seemed reasonable for the model to be confused (Appendix A Fig.~5).
Without seeing the entire instance, it was not possible to tell whether the model was actually wrong or the instances had incorrect ground-truth labels.
In other cases, the token snippets did not show enough context for the participant to fully understand the concept represented by the feature.
A feature's activations on a snippet of tokens can be influenced by other tokens in the instance that are not included in the snippet.
For example, if the user copies a snippet of tokens from the examples into the ``Test Feature'' input~(\cref{fig:saefarer_detail_sports}B), the activations on their tokens may not match the example, making it difficult to reproduce the behavior.
This also highlights a need for better ways for users to evaluate their understanding of what a feature represents and when it activates.

\textbf{Analyzing multiple features.}
SAEfarer's focus is on analyzing features independently.
The participants expressed interest in analyzing multiple related features at the same time.
For example, most participants were interested in seeing all the features that activate on a given piece of text, such as an existing instance or their own example.
P1 was curious about finding groups of related features and analyzing when they did and did not activate together.
P3 was interested in exploring features that were often activated together.
P5 wanted to extend that idea and identify groups of features that activate together and correlate with the model's predictions and errors.
P1, P2, and P3 were all interested in being able to search for features based on the tokens they tend to activate on.

\textbf{SAE quality.}
Some participants raised concerns about the quality of the SAEs they were analyzing.
This was in part due to features that were not interpretable or whose behavior was difficult to replicate with test cases.
For example, P2 stated that ``it's a nice tool, for sure, but it will be limited by the quality of the SAEs'', which they felt were not very good.
To improve our SAEs and reduce the number of features that lack clear interpretations, we could more robustly experiment with the size and sparsity of the SAEs.
We could also use more recent advancements in SAE architecture and training~\cite{rajamanoharan2024JumpingAheadImproving, bussmann2024BatchTopKSparseAutoencoders, bussmann2025LearningMultiLevelFeatures, fel2025ArchetypalSAEAdaptive}.
Additionally, we can make improvements to the SAE analysis process, such as ignoring padding tokens.

\section{Limitations and Future Work}

There are several limitations with our pilot evaluation, such as having only five participants who were all Ph.D. students.
Although they had relevant background knowledge, their approaches to exploring the model may not reflect those of professional ML practitioners who work with text classification models.
We did not assess how practitioners would use the tool over extended periods of time to analyze their own datasets and models.
We also did not evaluate SAEfarer with wider audiences that lack familiarity with SAEs.

SAEfarer's feature rankings are calculated based on the confusion matrices of the instances that cause each feature to activate.
One limitation with this is that the rankings only consider whether or not a feature activates for a given instance, but not the actual activation value.
A feature's low activations tend to be noisier and less important than its high activations ~\cite{bricken2023MonosemanticityDecomposingLanguage, templeton2024ScalingMonosemanticityExtracting}.
We may want to create additional rankings that give more weight to higher-value activations rather than treating all activations as equivalent.

Goyal et al.~\cite{goyal2020ExplainingClassifiersCausal} note that one limitation with approaches like TCAV is that they identify concepts that are correlated with the model's predictions, but not necessarily causal.
Our feature rankings suffer from a similar limitation in that they only indicate if a feature's activations are correlated with the model's predictions.
They do not indicate whether the concept has a causal impact on the model's predictions.

\section*{Supplemental Materials}
\label{sec:supplemental_materials}

Code and example notebooks for SAEfarer are available at \href{https://github.com/DanielKerrigan/saefarer}{github.com/DanielKerrigan/saefarer}.
Pilot evaluation material is available at \href{https://github.com/DanielKerrigan/saefarer-user-study}{github.com/DanielKerrigan/saefarer-user-study}.
We also include a demo video and appendices that provide screenshots of the UI, feature examples, and System Usability Scale results from the pilot evaluation.

\bibliographystyle{abbrv-doi}
\bibliography{ms}

\appendix
\onecolumn

\section{User Interface}
\label{appendix:ui}

\begin{figure*}[h]
    \centering
    \includegraphics[width=0.85\linewidth]{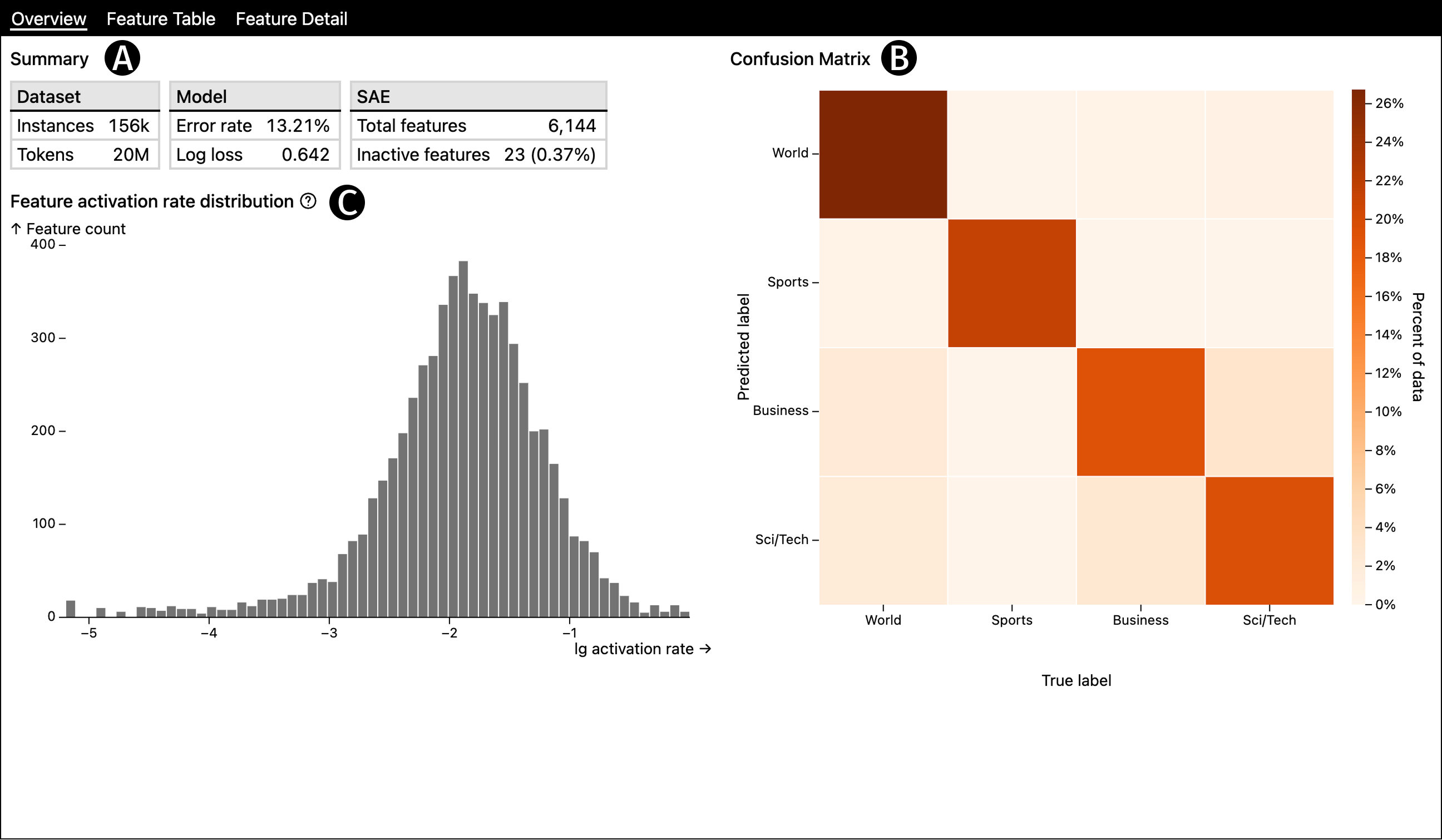}
    \caption{%
        The Overview tab in SAEfarer.
        A) Summary information about the dataset used to analyze the SAE, the classification model's performance, and the SAE.
        B) A confusion matrix showing the model's performance on all instances.
        C) A histogram showing the distribution of how often features in the SAE activate.
        The histogram plots the $\log_{10}$ activation rates.
    }
  \label{fig:saefarer_overview}
\end{figure*}

\begin{figure*}[h!]
    \centering
    \includegraphics[width=0.85\linewidth]{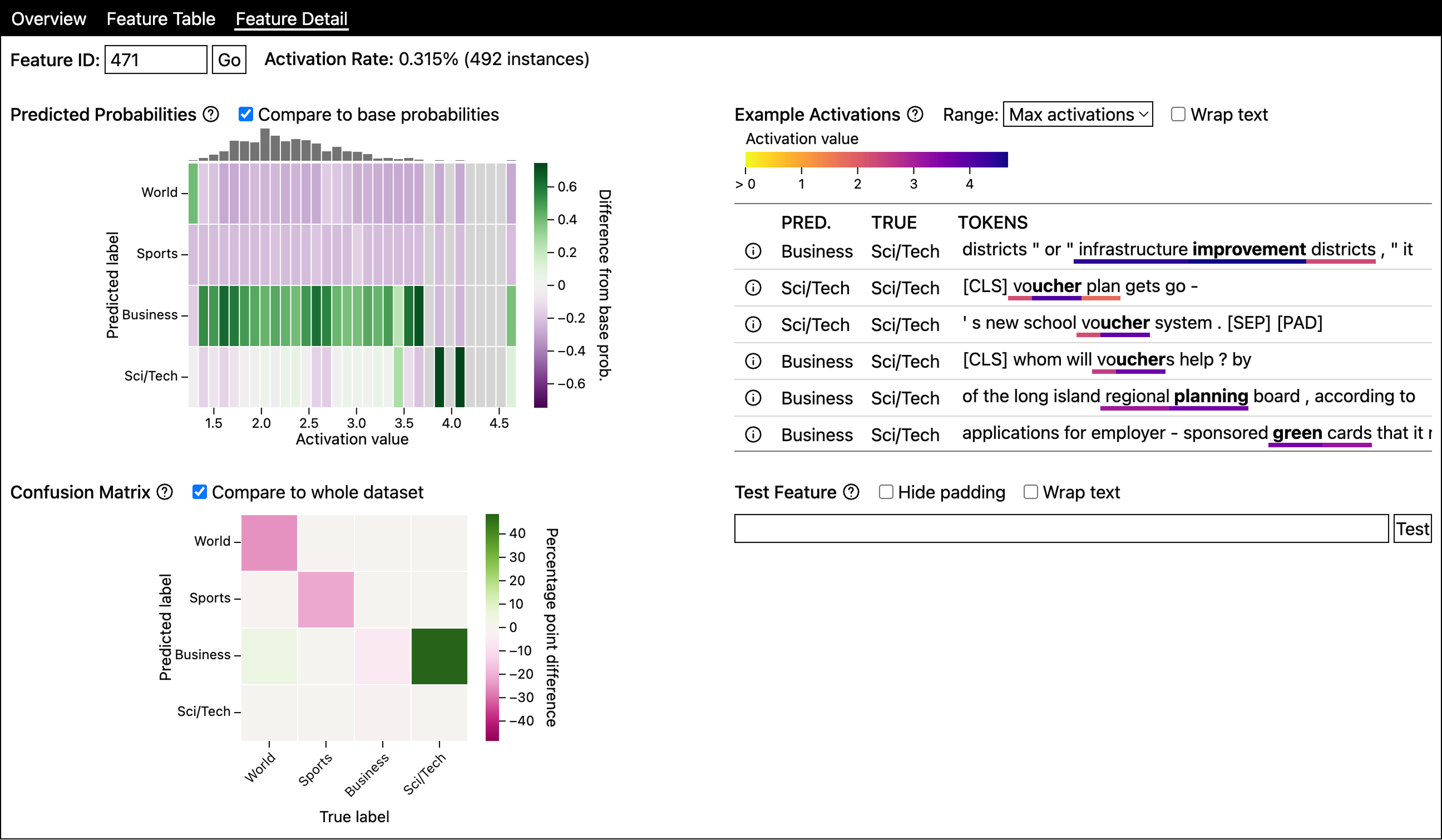}
    \caption{%
        The Feature Detail tab in SAEfarer for a feature that correlates with the model incorrectly classifying \textit{Sci/Tech} articles as \textit{Business}.
        Both the confusion matrix and marginal plot are colored to highlight the difference between the model's predictions on the instances that activated this feature vs. the model's predictions on the whole dataset. 
    }
    \label{fig:saefarer_detail_local_gov}
\end{figure*}

\begin{figure*}[h!]
    \centering
    \includegraphics[width=\linewidth]{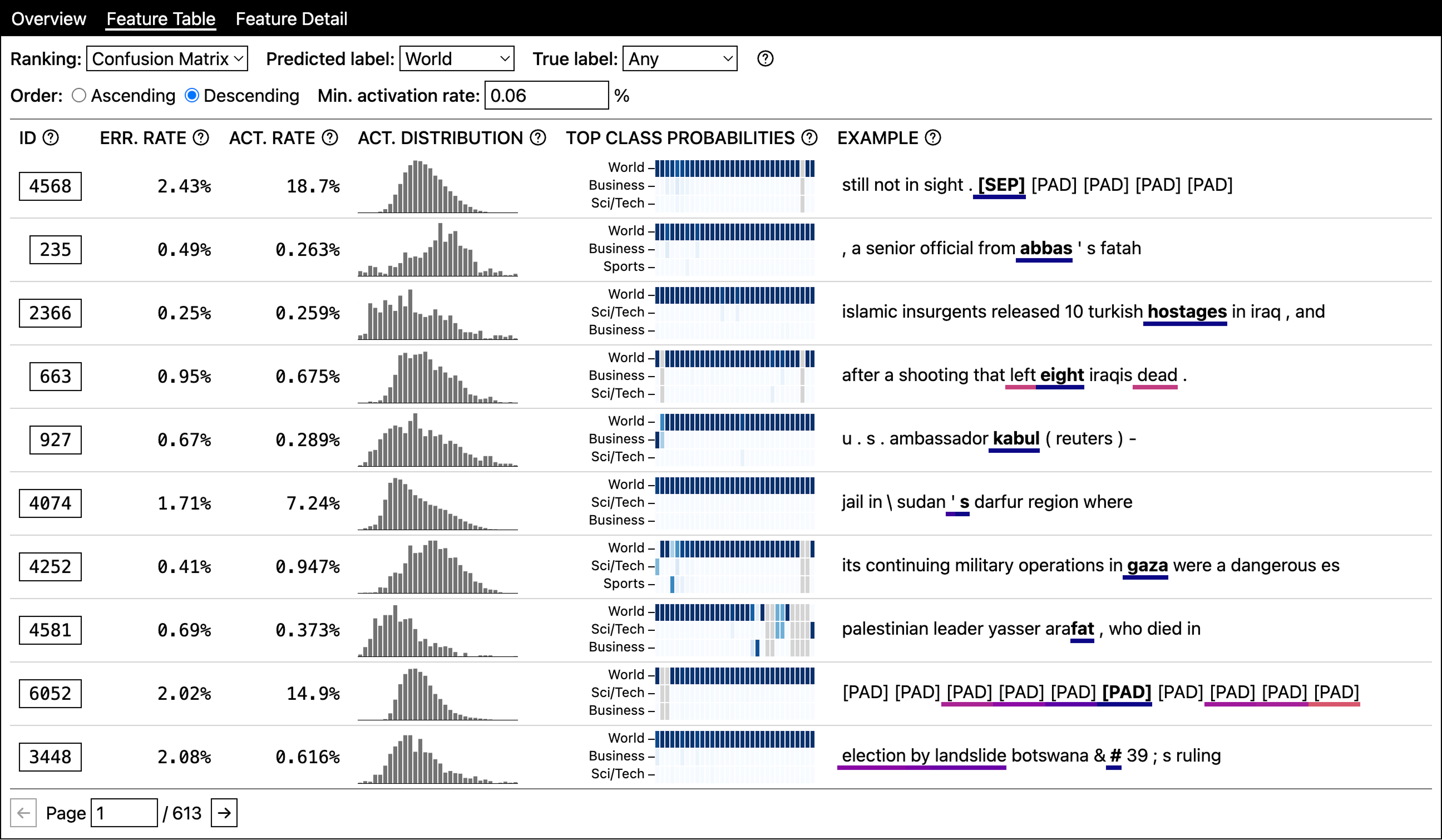}
    \caption{%
        The Feature Table tab in SAEfarer. Here, the features are ranked by their relationship to the model predicting the \textit{World} class.
        This table includes some examples of features that are spurious or difficult to interpret.
    }
    \label{fig:saefarer_table_world}
\end{figure*}

\clearpage

\section{Example Features}

\begin{figure*}[h]
    \centering
    \includegraphics[width=0.95\linewidth]{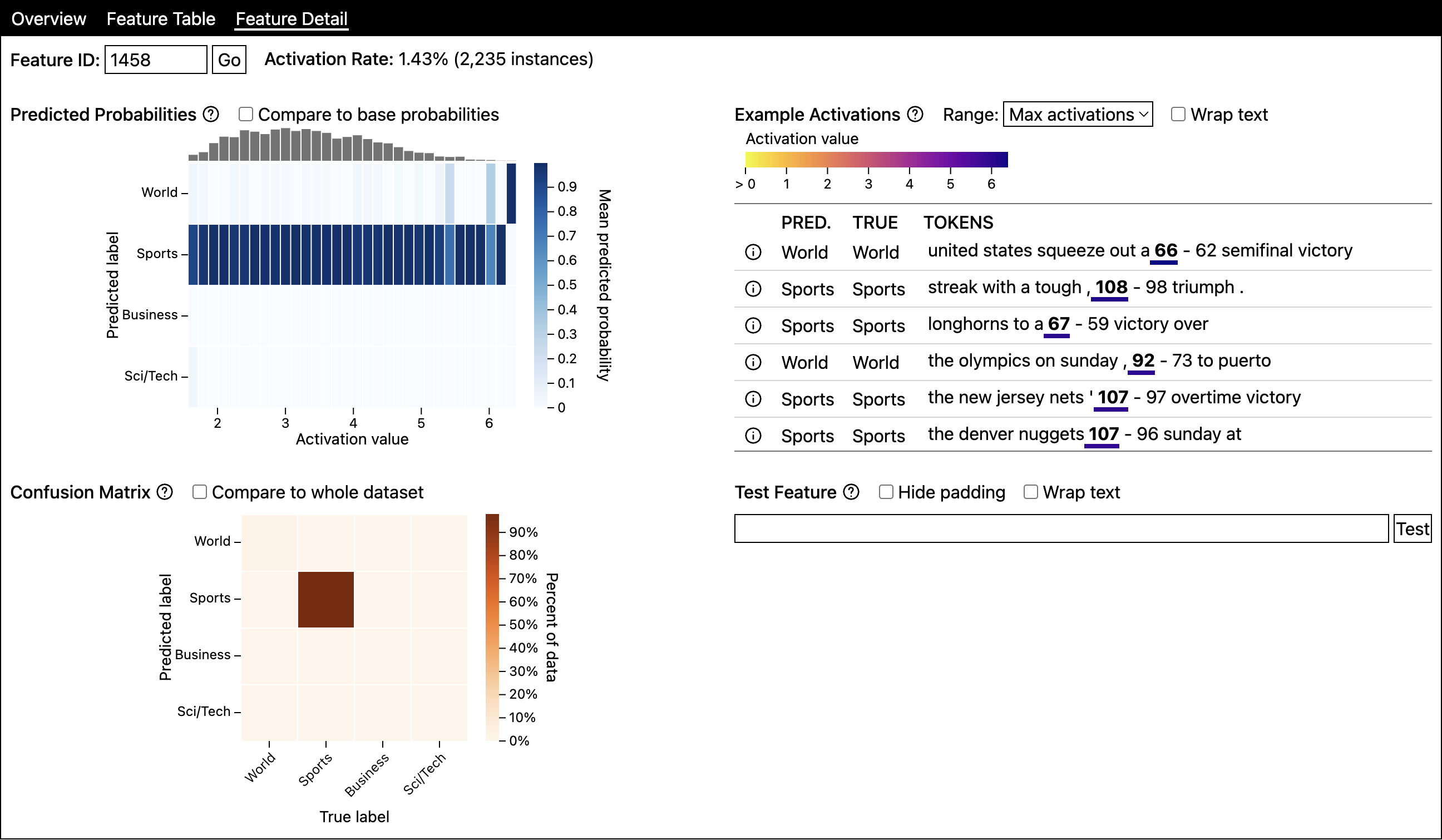}
    \caption[SAEfarer example feature: Sports scores.]{%
        This feature activated on the first number of scores for a sporting event, such as a basketball or football game.
    }
    \label{fig:saefarer_scores}
\end{figure*}

\begin{figure*}[h!]
    \centering
    \includegraphics[width=0.95\linewidth]{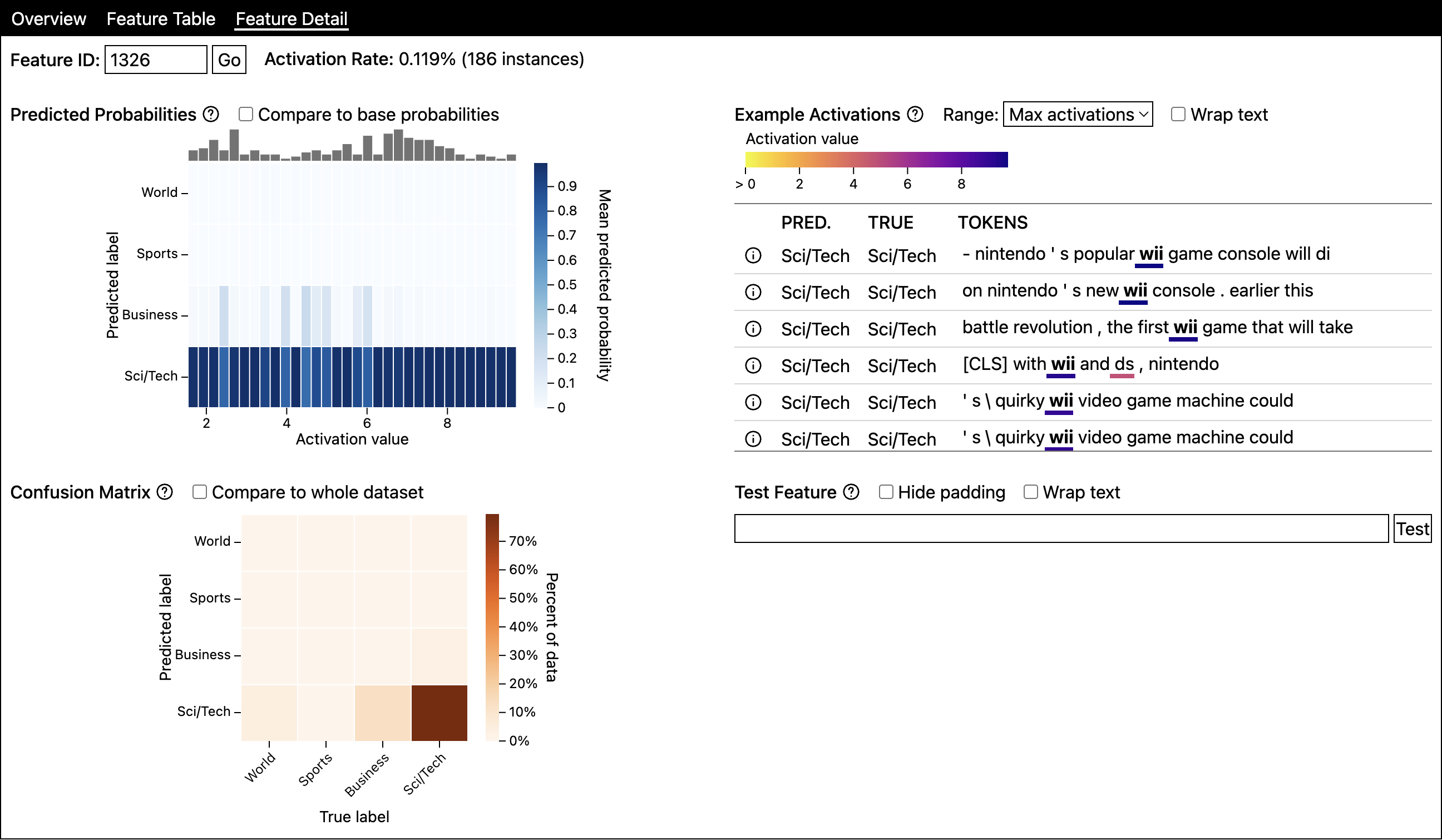}
    \caption[SAEfarer example feature: Nintendo Wii \& DS.]{%
        This feature activated on mentions of Nintendo Wii or Nintendo DS, which are two video game systems.
    }
    \label{fig:saefarer_wii}
\end{figure*}

\begin{figure*}
    \centering
    \includegraphics[width=0.95\linewidth]{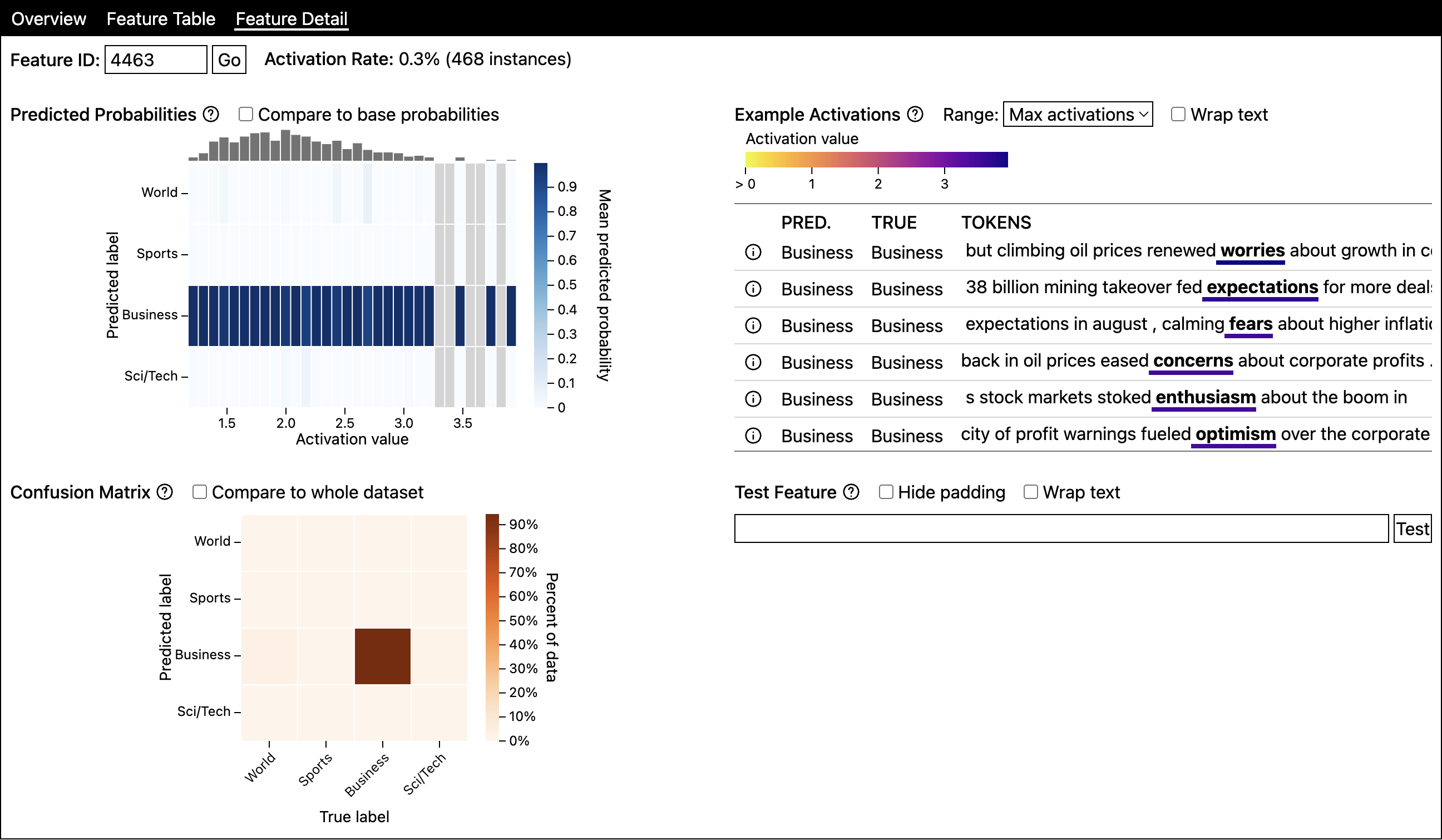}
    \caption[SAEfarer example feature: Emotions about the economy.]{%
        This feature activated on emotions relating to finance and economics.
    }
    \label{fig:saefarer_economy}
\end{figure*}

\begin{figure*}
    \centering
    \includegraphics[width=0.95\linewidth]{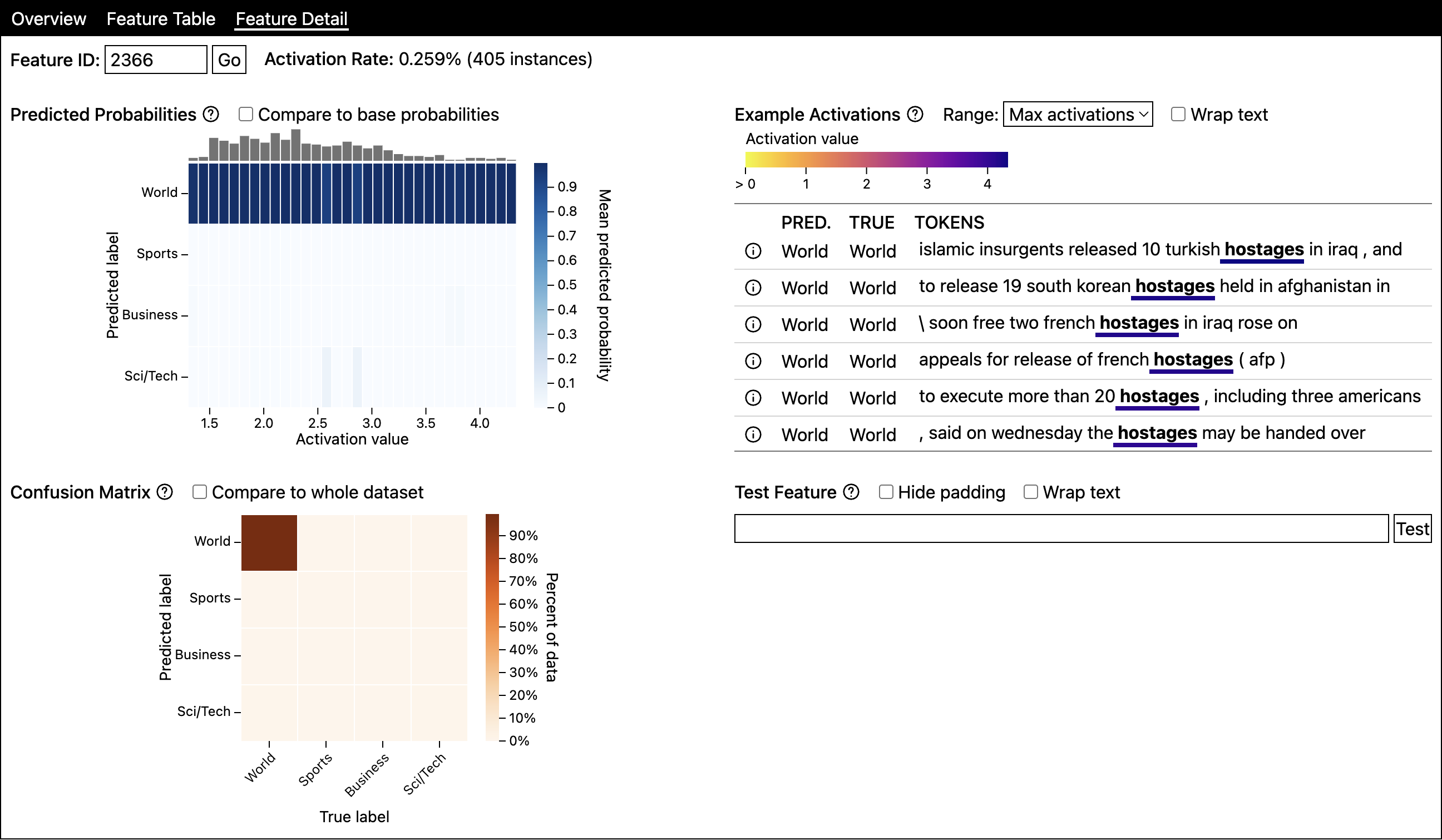}
    \caption[SAEfarer example feature: Hostages.]{%
        This feature activated on discussion of hostages.
    }
    \label{fig:saefarer_hostages}
\end{figure*}

\clearpage

\section{High vs. Low Activations}

\begin{figure*}[h]
    \centering
    \includegraphics[width=\linewidth]{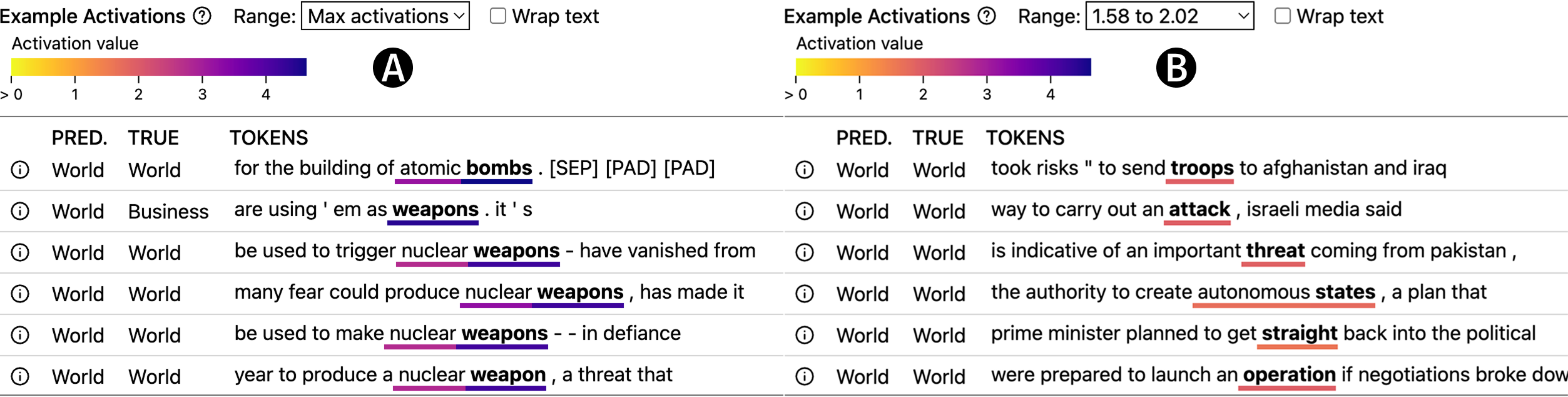}
    \caption[SAEfarer example feature: Nuclear weapons.]{%
        It is common for the maximum activations of a feature to have a clear focus and the lower activations to be noisier and less monosemantic.
        A) The maximum activations for a feature.
        These activations relate to nuclear weapons.
        B) Lower activations for the same feature.
        These activations seem related to international conflict more broadly.
    }
    \label{fig:saefarer_nukes}
\end{figure*}

\begin{figure*}[h]
    \centering
    \includegraphics[width=\linewidth]{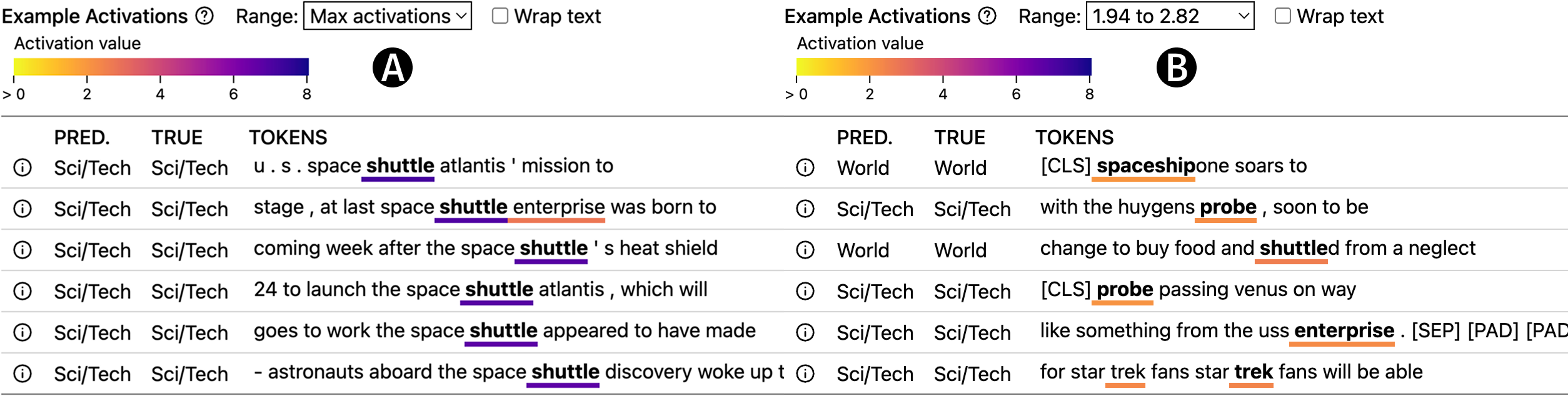}
    \caption[SAEfarer example feature: Space shuttle.]{%
        A) The maximum activations for a feature.
        These activations relate to the Space Shuttle.
        B) Lower activations for the same feature.
        Some of the activations are still related to spacecraft, while others are not.
        For example, in the third instance, the feature activates on ``shuttle'' in a context not about space.
        In the fifth instance, the feature activates on ``enterprise'' in the context of the USS Enterprise, which is an aircraft carrier, rather than the Space Shuttle Enterprise.
    }
    \label{fig:saefarer_shuttle}
\end{figure*}

\clearpage

\twocolumn

\section{System Usability Scale}

\begin{figure}
    \centering
    \includegraphics[width=\linewidth]{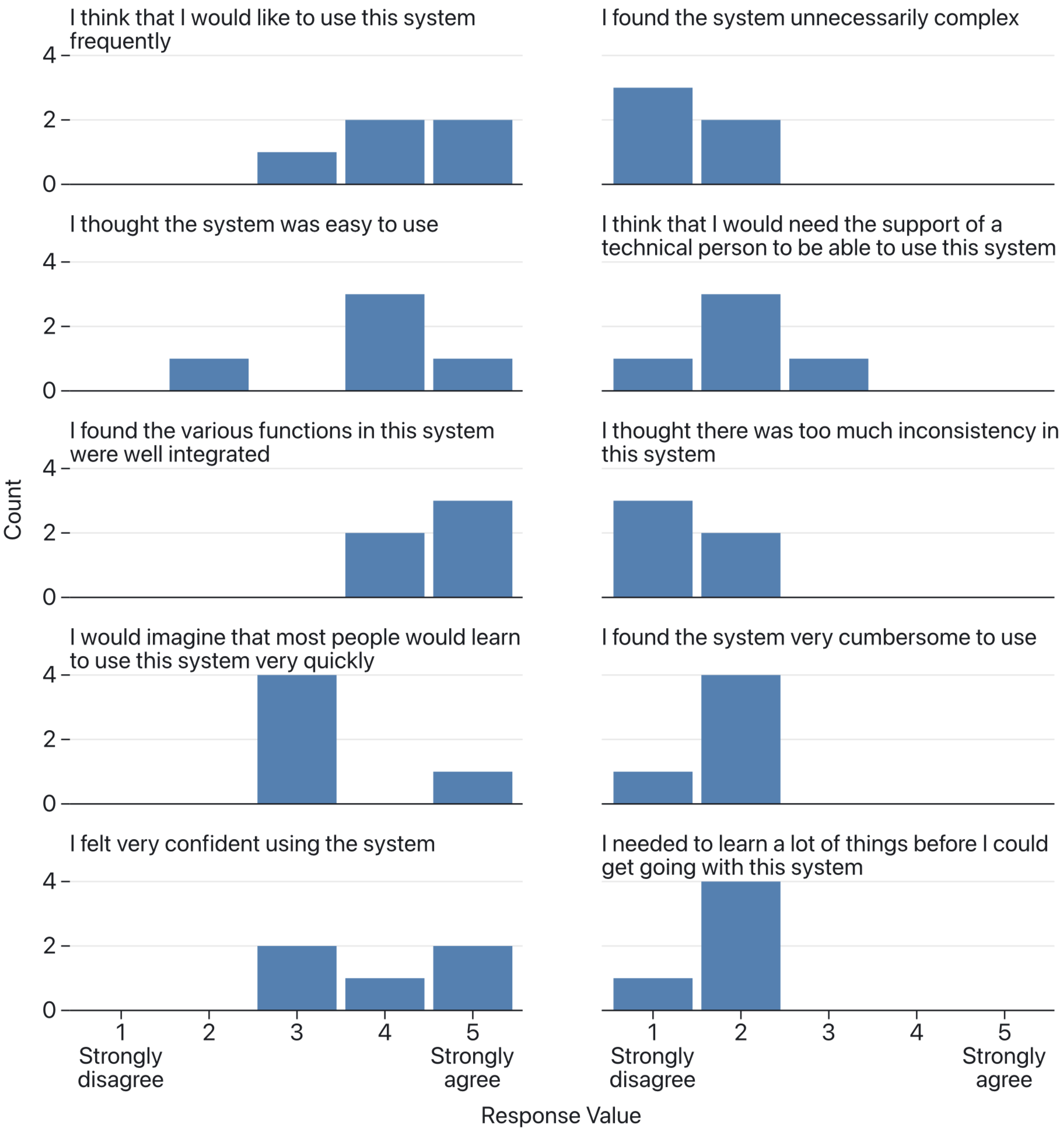}
    \caption[SAEfarer System Usability Scale results.]{%
        The results of the System Usability Scale. There is one bar chart for each item in the questionnaire. Each bar chart shows the number of participants that responded to the item with a given answer.
    }
    \label{fig:saefarer_sus}
\end{figure}

In the pilot evaluation, after participants completed analyzing the model and before we interviewed them, we had them fill out the System Usability Scale (SUS)~\citeapp{brooke1996SUSQuickDirty}.
The SUS consists of ten Likert items that ask the participant about their experience using the system.
The SUS score aggregates a participant's responses to the ten items in the questionnaire on a scale of 0 to 100.
To help interpret the SUS scores, they can be mapped onto a seven-point adjective scale with the values ``Worst Imaginable'', ``Awful'', ``Poor'', ``OK'', ``Good'', ``Excellent'', and ``Best Imaginable''~\citeapp{bangor2009DeterminingWhatIndividual}.
The participants' SUS scores were 90, 70, 80, 80, and 75, respectively, which averages to a score of 79.
On the adjective scale, a score of 79 maps to ``Good''.
\Cref{fig:saefarer_sus} shows the participants' responses to each item.

\crefalias{section}{appendix}
\bibliographystyleapp{abbrv-doi}
\bibliographyapp{ms}

\end{document}